\documentclass[pra,prl,twocolumn,superscriptaddress,showpacs,floatfix,10pt]{revtex4}
\usepackage{mathrsfs}
\usepackage{amsfonts}
\usepackage{amsmath}
\usepackage{amssymb}
\usepackage{graphicx}
\usepackage{footmisc}
\usepackage{braket}
\begin{document}

\title{Filling the gap between quantum no-cloning and classical duplication}
\author{Ming-hao Wang}
\affiliation{State Key Laboratory of Magnetic Resonances and Atomic and Molecular Physics,
Wuhan Institute of Physics and Mathematics, Chinese Academy of Sciences, Wuhan 430071, China}
\affiliation{University of Chinese Academy of Sciences, Beijing 100049, China}
\author{Qing-yu Cai}
\thanks{Corresponding author. \\Electronic address: qycai@wipm.ac.cn.}
\affiliation{State Key Laboratory of Magnetic Resonances and Atomic and Molecular Physics,
Wuhan Institute of Physics and Mathematics, Chinese Academy of Sciences, Wuhan 430071, China}
\begin{abstract}
The correspondence principle suggests that a quantum description for the microworld should be naturally transited to a classical description within the classical limit. However, it seems that there is a large gap between quantum no-cloning and classical duplication. In this paper, we prove that a classical duplication process can be realized using a universal quantum cloning machine. In the classical world, information is encoded in a large number of quantum states instead of one quantum state. When tolerable errors occur in a small number of the quantum states, the fidelity of duplicated copies of classical information can approach unity. That is, classical information duplication is equivalent to a redundant quantum cloning process with self-correcting.
\end{abstract}
\maketitle

\section{Introduction}
It is well known that one cannot copy an unknown quantum state precisely~\cite{wootters1982single,dieks1982communication}.
Following the no-cloning theorem for pure states, the possibility of cloning a mixed state is also ruled out~\cite{barnum1996noncommuting}.
What we can do for an unknown quantum state is either obtain imperfect copies each time~\cite{buvzek1996quantum}, or obtain perfect copies with nonzero probability of failure ~\cite{duan1998probabilistic}.

In recent years, much progress has been made in studying quantum cloning machines, both theoretically and experimentally~\cite{scarani2005quantum,Fan2014Quantum}. Various kinds of cloning machines, including universal quantum cloning machines~\cite{gisin1997optimal,bruss1998optimal,werner1998optimal}, phase-covariant cloning machines~\cite{bruss2000phase,karimipour2002generation}, asymmetric quantum cloning machines~\cite{cerf2000pauli,Iblisdir2005Generalised}, and probabilistic quantum cloning machines~\cite{duan1998probabilistic} have been proposed. Furthermore, continuous variable cloning, has also been studied~\cite{Cerf2000Optimal,Braunstein2000Optimal}.

In the classical world, one can duplicate information precisely, generating arbitrary copies, e.g., we can copy a file on a computer or obtain many copies of a newspaper or a textbook, except in the case of statistical ensembles~\cite{daffertshofer2002classical}. For quantum theory to become the most general theory in the world, we should fill in the gap between quantum no-cloning and classical duplication. In other words, we should explain the reason why we can duplicate classical information precisely under the rule of quantum no-cloning.
In fact, Bohr's correspondence principle requires that there should be a natural transition from quantum no-cloning to classical duplication; however, such a transition has not been found in the more than three decades that have passed since the discovery of quantum no-cloning~\cite{wootters1982single,dieks1982communication}

In this paper, we show that a classical bit is not encoded in a single quantum state but rather in a large number of almost-identical quantum states. We calculate the fidelity of classical information using a universal quantum cloning machine. When the number of almost-identical quantum states is sufficiently large and the errors that occur in a small number of theses states can be tolerated in the quantum clone process, the fidelity of the classical bit, i.e., the information fidelity, can approach unity. This implies that a perfect duplication process for classical information can be realized under the rule of quantum no-cloning.

\section{Information and quantum no-cloning}
\label{sec:quantum-no-cloning}
The term ``information" has different meanings in different contexts, e.g., communication, knowledge, and reference.
In Shannon's theory, information is used to eliminate random uncertainty \cite{shannon2001mathematical}.
Here, by quantum information, we means that information is encoded in a quantum state that follows the laws of quantum mechanics.
Usually, we use the qubit as the basic unit of quantum information. A qubit is a two-state quantum system that, unlike the conventional bit, can be a continuum of possible states as specified by its wavefunction:
\begin{equation}
\ket{\psi} = \alpha \ket{0}+\beta \ket{1}.
\end{equation}
Here, $\alpha$ and $\beta$ are arbitrary complex numbers, apart from the normalization condition. The fact that a particle could be in a superposition state keeps the state of an particle from being perfectly cloned~\cite{wootters1982single,dieks1982communication}, which may be the biggest difference between qubits and bits.
Unlike bits, which can be strictly distinguished, the superposition also makes states of a qubit undistinguishable~\cite{barnett2009quantum,chefles2000quantum}.
Currently, most scientists believe that quantum mechanics is the most precise theory for describing the world. The quantum description of a physical system can be naturally transited to classical description with the classical limit, which requires explaining why one can duplicate classical information under the rule of quantum no-cloning.

\section{Classical information and ensemble}
\label{sec:classical-information}
In the macroscopic world, a classical bit of $0$ or $1$ consists of numerous particles, the states of which are unknown.
We normally use discrete physical quantities to represent information, e.g., a two-state system, such as voltage, is often used to represent a bit. Although voltage is a continuous parameter, two well-separated regions can be chosen to represent $0$ and $1$.
Here, we give a brief summary of the properties of classical bits:
\begin{enumerate}
\item A classical bit consists of a large number of particles whose states are unknown for the duplication.
    The behavior of particles obeys laws of quantum mechanics.
\item States for different bits are well separated and can be entirely recognized.
\end{enumerate}
Since a classical bit consists of a large number of particles, it is better to describe a classical bit using the ensemble theory of quantum mechanics. Suppose that a classical bit is encoded in $N$ particles and that $N$ is a very large number. We use states of an ensemble of $N$ particles to denote the state of a classical bit. Given the $N$ particles in states $\ket{\psi_n}$ ($n=1,2,3...N$), we have the following density matrix
\begin{equation}
\rho=\frac{1}{N}\sum_{n=1}^{N}\ket{\psi_n}\bra{\psi_n},
\end{equation}
and the expectation of an observable quantity $\Omega$  is
\begin{equation}
\langle\Omega\rangle = Tr(\rho \Omega ).
\end{equation}
Regarding the environment, the states of the $N$ particles may change slightly as as their environment changes.
If the states of a small part, i.e., $\varepsilon$, are corrupted or even missed, i.e., $\ket{\psi_{n_i}}$ becomes $\ket{\psi_{n_i}'}$, the density matrix of the $N$ particles will become
\begin{equation}
\begin{aligned}\label{equ:rho}
\rho'=&\rho+\frac{1}{N}[\ket{\psi_{n_1}'}\bra{\psi_{n_1}'}+\ket{\psi_{n_2}'}\bra{\psi_{n_2}'}+...+\ket{\psi_{n_\varepsilon}'}\bra{\psi_{n_\varepsilon}'}\\
&-(\ket{\psi_{n_1}}\bra{\psi_{n_1}}+\ket{\psi_{n_2}}\bra{\psi_{n_2}}+...+\ket{\psi_{n_\varepsilon}}\bra{\psi_{n_\varepsilon}})],
\end{aligned}
\end{equation}
where $1\leq n_1<n_2<...<n_\varepsilon\leq N$.
When $\ket{\psi_{n_i}'}=\alpha_i\ket{\psi_{n_i}}+\beta_i\ket{\psi_{n_i}^{\bot}}$, $\alpha_i$ and $\beta_i$ are complex numbers satisfying the normalization condition, and $\ket{\psi_{n_i}^{\bot}}$ is orthogonal to $\ket{\psi_{n_i}}$, we can obtain
\begin{equation}
\begin{aligned}
\ket{\psi_{n_i}'}\bra{\psi_{n_i}'}= &\alpha_i\alpha_i^{\ast}\ket{\psi_{n_i}}\bra{\psi_{n_i}}+\beta_i\beta_i^{\ast}\ket{\psi_{n_i}^{\bot}}\bra{\psi_{n_i}^{\bot}}\\
&+\alpha_i^{\ast}\beta_i\ket{\psi_{n_i}^{\bot}}\bra{\psi_{n_i}}+\alpha_i\beta_i^{\ast}\ket{\psi_{n_i}}\bra{\psi_{n_i}^{\bot}}.
\end{aligned}
\end{equation}
The system of $N$ particles is normally embedded in a thermal environment; thus it will decohere quickly:
\begin{equation}\label{equ:decoherence}
\ket{\psi_{n_i}'}\bra{\psi_{n_i}'} \rightarrow \alpha_i\alpha_i^{\ast}\ket{\psi_{n_i}}\bra{\psi_{n_i}}+\beta_i\beta_i^{\ast}\ket{\psi_{n_i}^{\bot}}\bra{\psi_{n_i}^{\bot}}.
\end{equation}
Substituting Eq. (\ref{equ:decoherence}) into Eq. (\ref{equ:rho}) and simplifying, we can obtain the density matrix
\begin{equation}
\begin{aligned}
\rho'=&\rho+\frac{1}{N}[\ \beta_1\beta_1^{\ast}(\ket{\psi_{n_1}^{\bot}}\bra{\psi_{n_1}^{\bot}}-\ket{\psi_{n_1}}\bra{\psi_{n_1}})\\
&+\beta_2\beta_2^{\ast}(\ket{\psi_{n_2}^{\bot}}\bra{\psi_{n_2}^{\bot}}-\ket{\psi_{n_2}}\bra{\psi_{n_2}})...\\
&+\beta_\varepsilon\beta_\varepsilon^{\ast}(\ket{\psi_{n_\varepsilon}^{\bot}}\bra{\psi_{n_\varepsilon}^{\bot}}-\ket{\psi_{n_\varepsilon}}\bra{\psi_{n_\varepsilon}}) ].
\end{aligned}
\end{equation}
Hence, the average expectation will be
\begin{equation}
\begin{aligned}
\langle\Omega'\rangle =& Tr[\Omega\rho']=Tr[\Omega\rho]\\
&+\frac{1}{N}Tr[\ \Omega[\beta_1\beta_1^{\ast}(\ket{\psi_{n_1}^{\bot}}\bra{\psi_{n_1}^{\bot}}-\ket{\psi_{n_1}}\bra{\psi_{n_1}})\\
&+\beta_2\beta_2^{\ast}(\ket{\psi_{n_2}^{\bot}}\bra{\psi_{n_2}^{\bot}}-\ket{\psi_{n_2}}\bra{\psi_{n_2}})...\\
&+\beta_\varepsilon\beta_\varepsilon^{\ast}(\ket{\psi_{n_\varepsilon}^{\bot}}\bra{\psi_{n_\varepsilon}^{\bot}}-\ket{\psi_{n_\varepsilon}}\bra{\psi_{n_\varepsilon}})]\ ].
\end{aligned}
\end{equation}
The change in expectation is $\Delta=\mid\langle\Omega\rangle-\langle\Omega{'}\rangle\mid \leq \frac{2\varepsilon}{N} \cdot \lambda_{MAX}$,
where $\lambda_{MAX}$ denotes the maximum eigenvalue of $\Omega$. Therefore, for a large ensemble, some microcosmic finite errors do not
significantly change the macroscopical average of mechanical quantities and  $\langle\Omega'\rangle \approx Tr[\Omega\rho] =\langle\Omega\rangle $
since $\varepsilon$ is much less than $N$. Thus, we can use two ensembles to represent two classical bits $0$ or $1$.

\section{Classical information duplication}

\begin{figure}[htbp]
\centering\includegraphics[width=7cm]{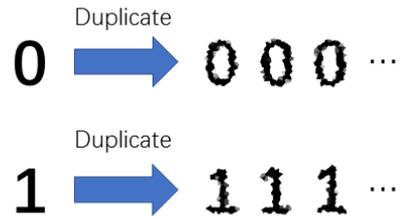}
\caption{An imperfect classical duplication. After duplication, we obtain several copies with some errors,
but the information can still be well recognized. Actually, the number of errors tha occur in a practical duplication process is
much less than that shown in this figure.}
\label{FIG1}
\end{figure}

In fact, the entire duplication process may be imperfect, and some small errors may occur in the duplication process due to the quantum no-cloning. However, even if some errors occur in the duplication process, the information for all copies is sufficiently to be distinguished.
As shown in FIG.~\ref{FIG1}, we can recognize $0$ or $1$ as original state despite the fact that errors occur.
We first consider the simple model in which classical information is encoded on an ensemble that contains $N$ particles
in identical quantum states, i.e., the state to be cloned is in $\ket{\psi}^{\bigotimes N}$. The practical case in which the states of
the $N$ particles are not identical will be discussed later.
Since there are no special requirements regarding output, we consider only the symmetric quantum cloning machine
for simplicity~\cite{gisin1997optimal}, and the $N$ particles are simplified to $N$ qubits.
Generally, the output state of the symmetric cloning machine is
\begin{equation}
\begin{aligned}
\ket{\psi_{out}}=&\sum_{j=0}^{M-N}\alpha_j\ket{(M-j)\psi,j\psi^\bot}\otimes R_j(\psi),\\
\alpha_j=&\sqrt{\frac{N+1}{M+1}}\sqrt{\frac{(M-N)!(M-j)!}{(M-N-j)!M!}},
\end{aligned}
\end{equation}
where $\ket{N\psi}$ denotes $N$ identical input states, $\ket{(M-j)\psi,j\psi^\bot}$ denotes  symmetric composite states
in which there are $M-j$ qubits in the state $\psi$ and $j$ qubits in the orthogonal state $\psi^\bot$, $R$ denotes the
 initial state of the copy machine, and $R_j(\psi)$ are orthogonal normalized internal states of the quantum cloning machine (QCM).

The reduced density matrix of $n$ qubits as a unit for the output is
\begin{equation}
\begin{aligned}
\rho_n = &\sum_{k=0}^{n}\sum_{j=k}^{Min(M-n+k,M-N)}\alpha_j\alpha_j^\ast \frac{C_{M-n}^{j-k}C_{n}^k}{C_M^j}\\
&\times\ket{(n-k)\psi,k\psi^\bot}\bra{(n-k)\psi,k\psi^\bot}.
\end{aligned}
\end{equation}
It is worth noting that here the value of $j$ ranges from $k$ to $M-n+k$, which may be greater than $M-N$, thus making $\alpha_j$ meaningless. We need to take the minimum between $M-n+k$ and $M-N$.
Specific to classical duplication, $M$ should be an integral multiple of $N$. We use $\kappa$ as a proportionality coefficient, that is, $M= \kappa N$, with $\kappa$ being much less than N. Moreover, when cloning finished, we obtain $\kappa$ copies of the input information since the output states of the $N$ qubits are duplications of the original state of classical information.
The key question becomes whether these $N$ qubits are close enough to the input $N$ qubits or not.
The QCM gives
\begin{equation}
\begin{aligned}
\rho_N = &\sum_{k=0}^{N}\sum_{j=k}^{\kappa N-N}\alpha_j\alpha_j^\ast \frac{C_{\kappa N-N}^{j-k}C_{N}^k}{C_{\kappa N}^j}\\
&\times\ket{(N-k)\psi,k\psi^\bot}\bra{(N-k)\psi,k\psi^\bot}.
\end{aligned}
\end{equation}
The corresponding fidelity is
\begin{equation}
\begin{aligned}
\emph{F}_{N,\kappa N}^{N}=&\sum_{j=0}^{\kappa N-N}\alpha_j\alpha_j^\ast \frac{C_{\kappa N-N}^{j}}{C_{\kappa N}^j}\\
=&\sum _{j=0}^{\kappa N-N} \frac{N+1}{\kappa N+1}\bigg(\frac{(\kappa N-N)!(\kappa N-j)!}{(\kappa N-N-j)!(\kappa N)!}\bigg)^2.
\end{aligned}
\end{equation}
Here, $\emph{F}_{N,\kappa N}^{N}$ denotes the fidelity between $N$ output qubits and $N$ input qubits of QCM.
\begin{figure}[htbp]
\centering\includegraphics[width=10cm]{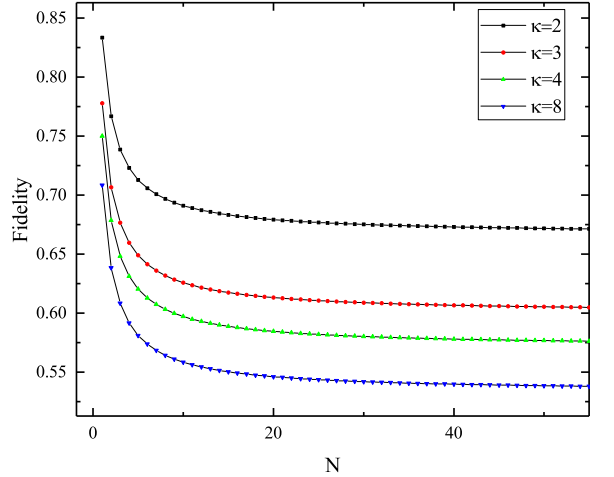}
\caption{(Color online) Here, $\kappa =2,3,4,8 $, which means that we obtain $2$, $3$, $4$ and $8$ copies. $\emph{F}_{N,\kappa N}^{N}$ decreases as $N$ increases.}
\label{FIG2}
\end{figure}

The numerical solution of $\emph{F}_{N,\kappa N}^{N}$ is shown in FIG.~\ref{FIG2}. The value of $\emph{F}_{N,\kappa N}^{N}$
decreases as the $N$ increases. This is counterintuitive, as we expect $\emph{F}_{N,\kappa N}^{N}$ approach 1 as $N$ increases. The reason why $\emph{F}_{N,\kappa N}^{N}$ decreases as $N$ increases is explained below.
In information theory, we normally use $000$ to replace $0$, whcih is the simplest way to protect the bits against the effects of noise.
Obviously, the probability of the output being $000$ is less than that of $0$. The reason why we can use $000$
against noise is that even if the state $000$ becomes $001$, $010$ or $100$, we can still decode it as $000$ ; thus, the information is kept. Similarly, as we have mentioned above, from the point of view of the macroscopical
average of a mechanical quantity, $\ket{(N-j)\psi,j\psi^\bot}$ is approximately equal to $\ket{N\psi}$ when $j$ is much less than $N$. In other words, if some errors occur in a small part of the code, we ignore the
errors and regard the code as either $0$ or $1$, as shown in FIG.~\ref{FIG1}. To describe this view precisely, we can
introduce a quantity named information fidelity,
\begin{equation}
\mathscr{F}= \sum_{\varepsilon=0}^{Err}\bra{\psi_\varepsilon} \rho \ket{\psi_\varepsilon}.
\end{equation}
Here, $\ket{\psi_{\varepsilon}}$ denotes a state with $\varepsilon$ particles having errors. $\varepsilon$ should be small enough
compared with the total number of particles $N$. The physical meaning of $\mathscr{F}$ is that the classical information
can still be kept even if states of $\varepsilon$ particles are erroneous, i.e., if $\varepsilon$ errors can be tolerated.
Using the definition above, we obtain the information fidelity of the output states for a symmetric quantum cloning machine as
\begin{equation}
\begin{aligned}
\mathscr{F}_{N,\kappa N}^{N}=&\sum_{\varepsilon=0}^{Err}\sum_{j=\varepsilon}^{\kappa N-N}\frac{N+1}{\kappa N+1}\frac{(\kappa N-N)!(\kappa N-j)!}{(\kappa N-N-j)!(\kappa N)!} \frac{C_{\kappa N-N}^{j-\varepsilon}C_N^\varepsilon}{C_{\kappa N}^j},
\end{aligned}
\end{equation}
where $Err$ represents the maximal number of errors that can be tolerated for an output state.

\begin{figure}[htbp]
\centering\includegraphics[width=10cm]{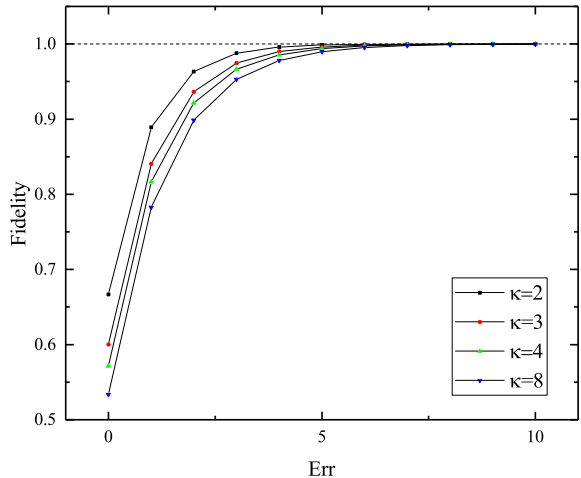}
\caption{(Color online) Numerical solutions of $\mathscr{F}_{N,\kappa N}^{N}$ with $N=1000$. $\mathscr{F}_{N,\kappa N}^{N}$ increases
and  rapidly approaches $1$ as $Err$ increases.}
\label{FIG3}
\end{figure}

The numerical solutions of $\mathscr{F}_{N,\kappa N}^{N}$ ($N=1000$) can be seen in FIG.~\ref{FIG3}.
When the parameters $\kappa$ and $N$ are fixed, $\mathscr{F}_{N,\kappa N}^{N}$ rapidly approaches $1$ as $Err$ increases.
Here, $Err$ ranges from $1$ to $10$, which is much less than $N$.
When $Err$ is greater than $5$, $\mathscr{F}$ is greater than $0.98$.
In practice, the minimal size for a classical bit may be at the nanoscale; otherwise the quantum effects will be notable.
For a nanoscale size, the number of particles is on the order of $10^{3}$, which indicates that the assumption od $N=1000$ is reasonable.
When $N$ is grater than $1000$, $10$ tolerated errors will cause the information fidelity to be sufficiently enough to $1$.
If $N$ continues to increase,the information fidelity will continually approach $1$ and the ratio $\frac{Err}{N}$ will approach to 0. The number of copies will also affect the quality of the duplication, though the effect can be ignored as $N$ increases.
$\mathscr{F}_{N,\kappa N}^{N}$ is a function of $\kappa$,$Err$ and $N$. Increases in the first two parameters will cause the function to monotonically decrease, while increase of $N$ will cause the function to monotonically increase.
Under macroscopic conditions, $N$ is much larger than $\kappa$ and $Err$, which makes a perfect duplication of classical information possible under the rule of quantum no-cloning.

\section{Discussion and conclusion}

In the calculation above, we assumed that $N$ particles for a classical bit are in an identical state $\ket{\psi}$ for simplicity.
In practice, some of these $N$ particles may be in states different from $\ket{\psi}$, i.e., $\ket{\Psi^{'}}$. Since $\ket{\Psi^{'}}$ is very similar to $\ket{\psi}^{\otimes N}$, we can express $|\Psi^{'}\rangle$ as
$|\Psi^{'}\rangle=\epsilon|\Psi\rangle+\sqrt{1-\epsilon^2}|\psi\rangle^{\otimes N}$. Here, $\epsilon$ is a small real number, and
$|\Psi\rangle$ is orthogonal to $|\psi\rangle^{\bigotimes N}$. It is obvious that the QCM is linear; thus, the cloning for $|\Psi^{'}\rangle$ can be
expanded into two parts: cloning $|\Psi^{'}\rangle$ and cloning $|\psi\rangle^{\otimes N}$. Since $\epsilon^2$ is a very small quantity, the cloning for $|\Psi^{'}\rangle$ is almost identical to that for $|\psi\rangle^{\otimes N}$.

Contrary to quantum no-cloning, which rules out the possibility of perfectly cloning an unknown quantum state, classical information can be duplicated perfectly, regardless whether the message is plaintext or ciphertext.
Since a classical bit is encoded in numerous micro particles instead of one micro particle, a classical duplication
is equal to a quantum cloning for an $N$-particle quantum state. When a small number of the error occur in the
$N$-particle state cloning can be tolerated or ignored, a classical duplication can be realized. That is, a classical
duplication process is equivalent to a redundant quantum cloning process with errors corrected based on the majority.
Although the $N$ particles are almost in the same quantum states, one does not need to know which quantum states they are in.
One may argue that the quantum states of the partial particles of a classical bit are quite different from those of others, and thus that our calculation may be invalid. In fact, very different states can be distinguished well
and can be used to represent different classical bits. In principle, the quantum states of particles at the minimal size for a classical bit should be
nearly identical.

\section*{Acknowledgments}

Finial support from National Natural Science Foundation of China under Grant Nos. 11725524, 61471356 and 11674089 is gratefully acknowledged.

\end{document}